\documentclass[]{elsarticle}

\usepackage{amssymb}
\usepackage{graphicx} 
\usepackage{textcomp}
\usepackage{times}
\usepackage{amssymb}
\usepackage{amsmath}
\usepackage{color} 
\usepackage{ulem}

\usepackage{lineno}

\newcommand{\ldot}{.\,}
\newcommand{\mylab}[1]{\label{#1}}

\graphicspath{{.}}%

\begin{document}

\begin{frontmatter}

%% Title, authors and addresses

%% use the tnoteref command within \title for footnotes;
%% use the tnotetext command for the associated footnote;
%% use the fnref command within \author or \address for footnotes;
%% use the fntext command for the associated footnote;
%% use the corref command within \author for corresponding author footnotes;
%% use the cortext command for the associated footnote;
%% use the ead command for the email address,
%% and the form \ead[url] for the home page:
%%
%% \title{Title\tnoteref{label1}}
%% \tnotetext[label1]{}
%% \author{Name\corref{cor1}\fnref{label2}}
%% \ead{email address}
%% \ead[url]{home page}
%% \fntext[label2]{}
%% \cortext[cor1]{}
%% \address{Address\fnref{label3}}
%% \fntext[label3]{}

%\dochead{}
%% Use \dochead if there is an article header, e.g. \dochead{Short communication}

\title{A constitutive law for cross-linked actin networks by homogenization techniques}

%% use optional labels to link authors explicitly to addresses:
%% \author[label1,label2]{<author name>}
%% \address[label1]{<address>}
%% \address[label2]{<address>}

\author[l1]{Denis Caillerie}
\address[l1]{L3S-R, BP 53 - 38041 Grenoble Cedex 9, France}
\ead{denis.caillerie@hmg.inpg.fr}

\author[l2]{Karin John}
\ead{karin.john@ujf-grenoble.fr}

\author[l2]{Chaouqi Misbah}
\ead{chaouqi.misbah@ujf-grenoble.fr}

\author[l2]{Philippe Peyla}

\address[l2]{LSP UMR 5588, Universit\'e J. Fourier and CNRS, BP 87 - 38402 Grenoble
Cedex, France}

\author[l3]{Annie Raoult}
\address[l3]{Laboratoire MAP5 UMR 8145, Universit\'e Paris Descartes,
45 rue des Saints P\`eres, 75270 Paris Cedex 06, France}

\begin{abstract}
Inspired by experiments on the actin driven propulsion of micrometer
  sized beads we develop and study a minimal mechanical model of a
  two-dimensional network of stiff elastic filaments grown from the
  surface of a cylinder.
  Starting out from a discrete model of the network structure and of
  its microscopic mechanical behavior we derive a macroscopic
  constitutive law by homogenization techniques.
  We calculate the axisymmetric equilibrium state and study its linear
  stability depending on the microscopic mechanical properties.
  We find that thin networks are linearly stable, whereas thick
    networks are unstable.  The critical thickness for the change in
    stability depends on the ratio of the microscopic elastic
    constants. The instability is induced by the increase in the
    compressive load on the inner network layers as the thickness of
    the network increases.

    The here employed homogenization
    approach combined with more elaborated microscopic models can
    serve as a basis to study the evolution of polymerizing actin
    networks and the mechanism of actin driven motion.
\end{abstract}

\begin{keyword}
A-microstructure \sep B-biological material \sep B-constitutive behavior \sep C-stability and bifurcation \sep homogenization
%% MSC codes here, in the form: \MSC code \sep code
%% or \MSC[2008] code \sep code (2000 is the default)
\MSC[2010] 74Q05 %Homogenization in equilibrium problems
\sep
74Q15 %Effective constitutive equations 
\end{keyword}

\end{frontmatter}

%%
%% Start line numbering here if you want
%%
%%\linenumbers

%% main text
\section{Introduction}
\mylab{intro}
Dynamic filamentous polymer networks play an essential role in the
mechanics of living cells.
However, adequate constitutive equations, which take into account the
growth history of the networks and the thereby developping
prestresses, are difficult to obtain.
One prominent example, which shall serve as a starting point for this
paper, is the actin filament network. 
Actin filaments cross-linked by a special protein complex called
Arp2/3 complex\footnote{The Arp2/3 complex is a special protein
  complex which induces the nucleation of new actin filaments at
  already existing ones, thus triggering the formation of
  Y-junctions. It is activated by other enzymes (e.g. ActA, Wasp)
  which are bound to the membrane or the surface of a biomimetic
  object and therefore Y-junction formation is limited to the space
  close to the enzyme covered surface.}  play a key role in cell
motility \cite{LBP99} as well as in the motion of cellular organelles
\cite{TRM00} or pathogens \cite{YTA99,CCG95} inside their respective
host cells.

This actin driven motion has been reconstituted under minimal
conditions \cite{LBP99} using beads \cite{OuT99}, vesicles
\cite{UCA03}, and droplets \cite{BCJ04}. The basic biochemical
processes necessary for motion are by now well characterized (for a
review see \cite{PBM00} and references therein).
Briefly, under the action of several auxiliary proteins, actin
polymerizes into a polar filament network, whereby filament elongation
and filament linkage are restricted to a zone close to the surface of
the propelled object, i.e. the internal interface (The filament
network is highly porous and does not hinder the diffusion of monomers
to the surface of the object.).
Depolymerization takes place far away from the object, i.e. at the
external interface. Therefore, new network material is inserted
between the already existing network and the object and forces the
older network away from the object [see Fig.\,\ref{growthscheme} (a)].
If this type of growth is occurring on curved surfaces, the network
experiences large stresses and deformation as growth proceeds.
\begin{figure}
\begin{center}
\includegraphics[width=0.7\hsize]{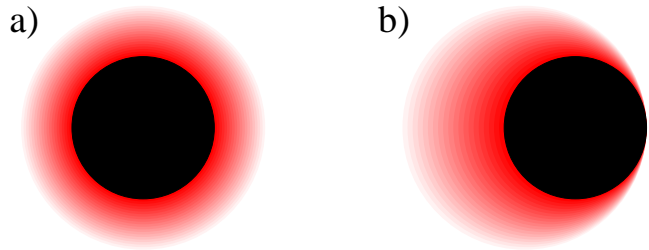}
\end{center}
\caption{The actin network (shades in red) grows initially in a
  symmetric cloud (a) from a bead surface (black) until a spontaneous
  symmetry breaking occurs and the growth continues in an asymmetric
  fashion (b). Newer (older) network layers are in dark (light)
  red.\mylab{growthscheme}}
\end{figure}
In experiments with actin-propelled biomimetic objects (e.g.~beads)
\cite{OuT99,UCA03,BCJ04} it was shown, that initially the actin
networks forms a symmetrical cloud around the object, until a
spontaneous symmetry-breaking occurs [see
Fig.\,\ref{growthscheme} (b)]. Then the network is growing faster on
one side than on the other and finally the object starts to move
pushed by an actin comet. Thereby actin is continuously polymerizing
in the contact region between the bead and the comet.

 Thus the (chemical) out-of-equilibrium process of
  polymerization/depolymerization (which is nevertheless stress
  dependent, see \cite{FKT07} and references therein) and the action
  of enzymes produce a network in a pre-stressed state, which does not
  relax on the timescale of symmetry-breaking and which depends on the
  history of growth \cite{OuT99,NGF00,GPP05,DSR08}. 
% 
%\cite{OuT99, MoO03,
%    GPP05,SPJ04,LLK05,NGF00,JPK08,JCP09}. 
%
  Several theoretical studies have addressed this problem either
  using discrete \cite{OuT99, MoO03} or phenomenological continuous models
  \cite{SPJ04,LLK05,JPK08} without, however, obtaining an adequate
  constitutive law for the network.
  A type of homogenization model has been employed previously to the
  actin network in the geometry of the lamellipodium
  \cite{OSS08}. However, this model has a viscoelastic character and
  stresses build up only transiently due to temporary cross-link
  formation and not due to static crosslinks formed close to the
  polymerizing interface of the network.
  The rigorous derivation of a macroscopic constitutive law of the
  network is therefore still lacking and is the subject of the present
  paper which takes up several ideas already outlined in \cite{JCP09}.
%
%
%The common notion is, that the elastic properties of the actin network
%are decisive for the symmetry-breaking \cite{OuT99, MoO03, GPP05,
%SPJ04,LLK05,NGF00,JPK08,JCP09}, since the growth process is stress
%dependent (see \cite{FKT07} and references therein).
%
%However, an adequate constitutive law of the network is still
%missing since deformations are large and the major source of stress is
%the growth of the network.
%

%
One crucial observation is that actin forms a more or less periodic
network on the microscopic scale, which is stable on the time scale of
symmetry-breaking (i.e. several minutes).
The size of each elementary cell, e.g.\ the distance between two links
in the actin network ($\sim$ several tens nm), is small compared to
the total size of the structure ($\sim$1\,\textmu{}m), which
introduces a small parameter $\eta$ into the problem and allows for
the upscaling of the microscopic mechanical model to a continuous
medium by discrete homogenization methods \cite{MoC98,ToC98,CMR03}.

In this paper we apply the discrete homogenization method to a simple
2D network structure which has been ``grown'' from a circle. While we
keep the network structure and the microscopic mechanical model
simple, we retain crucial biological features. In the {\bf Theory}
section we will define the network topology along with the microscopic
mechanical properties and derive a set of continuous equilibrium
equations.
In the {\bf Results} section we will solve these equations for the
axisymmetric equilibrium state and identify the conditions under which
linearly stable solutions exist. Finally, in the {\bf Conclusions}
section we will give some perspectives for this homogenization
approach in modeling actin networks.
\section{Theory} \mylab{model}
In this section we derive the continuous equilibrium equations
governing the homogenized network in an abstract Lagrangian coordinate
frame. We start out with the discrete formulation of the network
structure then upscale this description to a continuous medium.
In a second step we introduce a free energy for the continuous
medium and derive the equations of equilibrium using the first
variation of the free energy.
For completeness we outline in {\bf \ref{appa}} and {\bf
  \ref{appb}} the relation between the equilibrium description based
on the stress vectors in the abstract Lagrangian frame as it is
introduced in this {\bf Theory} section and the classical Cauchy stress
tensor.
\subsection{Description of the discrete network}
\mylab{setup}
We consider a planar network around a solid circle of radius
$R_0$. The network consists of stiff elastic filaments linked at their
ends by nodes. Its topology is shown in Fig.\ \ref{nwtop} on the left.
\begin{figure}
\begin{center}
\includegraphics[width=0.7\hsize]{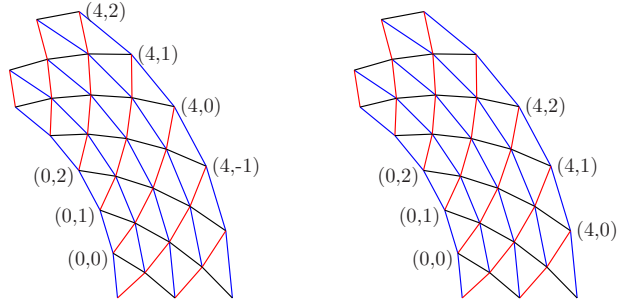}
\end{center}
\caption{[Color online] Sketch of a small part of the filament network
  with some examples for the node numbering before (left) and after
  (right) the transformation of variables $(\mu^1,\mu^2)\rightarrow
  (\lambda^1,\lambda^2)$ as introduced in Eqs.\,(\ref{cv1}) and
    (\ref{cv2}). The ``curves'' along which $\mu^1$=const. and
  $\mu^2$=const. are shown in blue and red, respectively.
  \mylab{nwtop}}
\end{figure}
The bars are attached to the cylinder surface at $N_{t}$ sites evenly
located at a distance $p=\eta R_0$, i.e.\ at an angular distance $\eta
=2\pi/N_t$.
The network is made of $N_{n}$ layers of bars in the radial direction.
As the network was formed by polymerization of actin monomers at the
surface of the cylinder, each radial layer contains the same number of
nodes $N_t$. Each node can be identified by two integers $\left({\nu
    ^{1},\nu ^{2}}\right)\in\{0,\ldots,N_n-1\}\times\{0,\ldots,N_t-1\}$ as shown in
Fig.\,\ref{nwtop} on the left.
It is assumed that $N_{t}$ and $N_{n}$ are very large and of the same
order. Consequently, $\eta= 2\pi/N_t$ is assumed to be very
small and the parameter $\alpha=\eta N_n$ is of order 1 with respect
to $\eta $.

The elementary cell of such a network consists of one node and three
bars $b=1,2,3$ and is shown in Fig.\,\ref{unitcell}. Orientation and
length of each bar is described by the vector $\vec B ^b$. The tension
in the bars is described by a linear law
\begin{equation}
N^{b}=k^{b}\frac{l^{b}-l_{m}^{b}}{l_{m}^{b}}\mylab{cl}
\end{equation}
with $l^{b}=\vert{\vert{\vec{B}^{b}}|}|$.
$k^b$ is a spring constant and $l_m^{b}$ is the length of the bar $b$
at rest. Here we will only consider the situation $k^{1}=k^{3}$ and
$l_{m}^{1}=l_{m}^{3}$.
At this stage the choice of a linear law is somewhat arbitrary and
constitutes the simplest possible choice. It is consistent with basic
principles in mechanics, such as material frame indifference. However,
the homogenization procedure is valid for any microscopic law.

So far we have only considered forces associated with the lengths
  of the filaments and have not included the fact, that we are only
  interested in networks which keep a certain topology, e.g.~the
  elementary cell has a finite area $\parallel\vec B^1 \wedge \vec B^3\parallel > 0$. To
  avoid this type of transition, i.e. a collapse of the network into
  zero area and a complete alignment of the vectors $\vec B^b$ we add
the following interaction moment between filaments 1 and 3
\begin{equation}
\vec M^{13}=M^{13}\vec B^1\wedge \vec B^3\mylab{mm}
\end{equation}
with
\begin{equation}
M^{13}=\kappa\frac{(l^1l^3)^2(\vec B^1\ldot \vec B^3)}{\left[(l^1l^3)^2-(\vec B^1\ldot \vec B^3)^2\right]^2} \mylab{mm1}
\end{equation}
where $\kappa$ is a constant and denotes the interaction strength. As
will be seen later in the derivation of the equilibrium equations
in section \ref{secee}, expressions (\ref{mm}) and (\ref{mm1}) arise
naturally from the first variation of the free energy which contains a
potential which diverges for the complete alignment of $\vec B^1$ and
$\vec B^3$.

%% %
%% The Cauchy stress tensor $\sigma$ at any given node in the network is
%% given by the classical relation (see appendix or \cite{Lov44,CaC01})
%% %
%% \begin{equation}
%% \sigma=\frac{1}{g}   \sum\limits_{\mit b = 1}^{3}\frac{N^{b}}{l^b}\vec{B}\,^{b}\otimes \vec{B}\,^{b}
%% \mylab{stress1}
%% \end{equation}
%% %
%% with $g=\left|\!\left|{\vec{B}}^{1}\wedge {\vec{\mit B}}^{\mit
%% 3}\right|\!\right|$ being the surface of the elementary cell.
%
\begin{figure}
\begin{center}
\includegraphics[width=0.7\hsize]{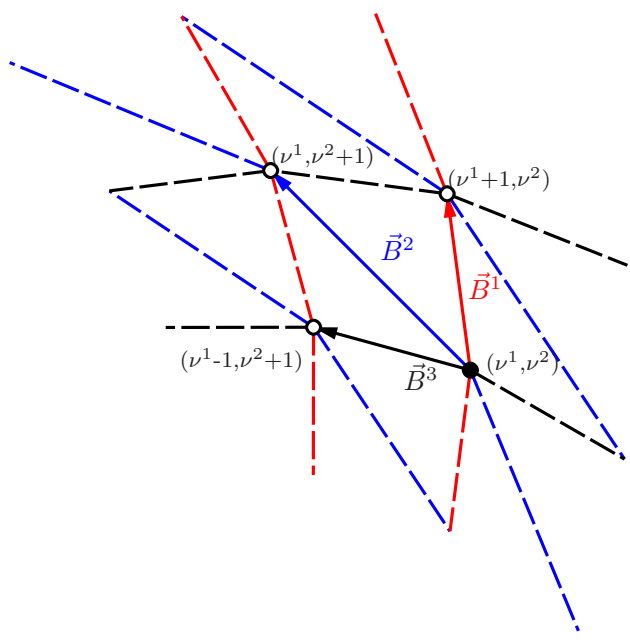}
\end{center}
\caption{[Color online] Sketch of an elementary cell, showing the
node numbering $(\nu^1,\nu^2)$ and the elementary bar vectors $\vec
B^b$ (solid arrows). All other bars are shown as dashed lines. The
``curves'' along which $\nu^1$=const. and $\nu^2$=const. are shown in
blue and red, respectively.\mylab{unitcell}}
\end{figure}

\subsection{Homogenization}
The upscaling of the network to a continuous medium consists in
determining the equivalent stresses from the bar tensions, the
equations of equilibrium (or motion) satisfied by these stresses and
an equivalent constitutive equation ensuing from the properties of
the bars.
This can be carried out by using an asymptotic expansion (for an
introduction see \cite{CaC01,Lov44,ToC98}). Here, as the network
structure is simple, a more heuristic presentation can be used.
The basic idea of the homogenization process is, that for most of the
network motions, the positions of its nodes $\left({\nu ^{1},\nu
    ^{2}}\right)$ can be approximated by a continuous deformation
function $\vec{\psi }\left({\mu ^{1},\mu ^{2}}\right)$ such that the
position of the node $(\nu^1,\nu^2)$ is $\vec{\psi }\left({\mu
    ^{1\eta},\mu ^{2\eta}}\right)$ with $\mu ^{i\eta}=\eta \nu ^{i}$
and $\left(\mu ^{1\eta },\mu ^{2\eta
  }\right)\in\{0,\eta,\ldots,\alpha-\eta\}\times\{0,\eta,\ldots,2\pi-\eta\}$.
The discrete Lagrangian variables $\left(\mu ^{1\eta },\mu ^{2\eta
  }\right)$ become then the set of Lagrangian (curvilinear)
coordinates $(\mu^1,\mu^2)\in \omega$ of the equivalent continuous
medium with $\omega =]0,\alpha [\times]0,2\pi [$.
%
%The purpose then is to determine the equations governing this
%deformation function $\vec{\psi }\left({\mu ^{1},\mu ^{2}}\right)$.
%

In this homogenized network the filament vectors $\vec B^b$ can now be
expressed using a simple Taylor expansion up to $\mathcal{O}(\eta)$
%
% vectors B
 \begin{eqnarray}
\vec{B}^{1}  & =& \eta \partial _{\mu^1}\vec{\psi }\nonumber\\
\vec{B}^{2} & = & \eta \partial _{\mu^2}\vec{\psi }\\
\vec{B}^{3} & = &\eta \left({-\partial _{\mu^1}\vec{\psi }+\partial _{\mu^2}\vec{\psi }}\right)
\end{eqnarray}
where $\partial _{\mu^i}=\frac{\partial }{\partial \mu ^{i}}$.
Since they are associated with a quite simple numbering system for the
nodes, the variables $\mu ^{1}$ and $\mu ^{2}$ arise naturally as
Lagrangian variables of the equivalent continuous medium through the
homogenization process. However, they are not the most convenient
variables to study axisymmetric equilibrium configurations. Therefore
we have introduced the variables $\lambda ^{1}$ and $\lambda ^{2}$
defined by
%
% Variable change
  \begin{equation}
\lambda ^{1}=\mu ^{1}\,\textrm{and}\,\lambda ^{2}=\frac{\mu ^{1}}{2}+\mu ^{2}
\mylab{cv1}
\end{equation}
and which correspond to the radial and angular position of the node in
the network [see Fig.\,\ref{nwtop} to the right].
Setting
% Def of phi
%
\begin{equation}
\vec{\varphi }\left({\lambda ^{1},\lambda ^{2}}\right)=\vec{\psi
}\left({\lambda ^{1},-\frac{\lambda ^{1}}{2}+\lambda ^{2}}\right)
\mylab{cv2}
\end{equation}
one finds
%
% vectors B
\begin{eqnarray}
\vec{B}^{1} & =& \eta \left(\vec{h}_{1}+\frac{1}{2}\vec h_2\right)\nonumber\\
\vec{B}^{2} & =& \eta \vec{h}_{2}\\
\vec{B}^{3} & =& \eta\left({-\vec{h}_{1}+\frac{1}{2}\vec{h}_{2}}\right)\nonumber
\end{eqnarray}
with $\vec{h}_{1}=\partial _{\lambda^1}\vec{\varphi }$ and
$\vec{h}_{2}=\partial _{\lambda^2}\vec{\varphi }$ where
$\partial_{\lambda^i}=\frac{\partial}{\partial_{\lambda^i}}$. In the
  following we will employ the simplified notation
  $\partial_i=\partial_{\lambda^i}$.

\subsection{Equations of equilibrium}
\mylab{secee}
The equations of equilibrium can be obtained directly by following the
homogenization procedure up to first order in $\eta$ detailed in
\cite{CMR03}.
However, here we have chosen a simpler approach using only the
filament vectors $\vec B^b$ obtained by homogenization and by defining
a free elastic energy $F$ of the network corresponding to the
constitutive functions $N^b$ and the moment $\mathbf{M}^{13}$ defined
in (\ref{cl}) and (\ref{mm}).
This free energy $F$ can be expressed as the functional
\begin{equation}
F[\vec\varphi] = \frac{1}{\eta^2}\int_\omega \left[\sum_{b=1}^3
\frac{k^b}{2l^b_m}\left(l^b-l^b_m\right)^2 + 
\frac{\kappa}{2}\frac{(\vec B^1\ldot \vec B^3)^2}{(l^1l^3)^2-(\vec B^1\ldot \vec B^3)^2}\right]d\lambda^1\,d\lambda^2\mylab{elfe}
\end{equation} 
where $l^b=\|\vec B^b\|$. The first sum in (\ref{elfe}) denotes the
free energy associated with the compression and elongation of the
filaments. The second term is associated with the interaction between
the filaments 1 and 3 and serves as a ``hard core'' potential, which
prevents the collapse of the networks, i.e. it prevents the complete
alignment of the filaments~1 and 3.

The first variation of $F$ with respect to the deformation function
$\vec \varphi$ reads
\begin{eqnarray}
\delta F & = &   \frac{1}{\eta^2}\int_\omega d\lambda^1\,d\lambda^2\left\{\sum_{b=1}^3 N^b \frac{\vec B^b}{l^b}\ldot\delta \vec B^b +\right.\nonumber\\ 
& & \left.\left(\vec M^{13}\wedge\frac{\vec B^1}{(l^1)^2}\right)\ldot\delta\vec B^1-\left(\vec M^{13}\wedge\frac{\vec B^3}{(l^3)^2}\right)\ldot\delta\vec B^3 \right\}\nonumber\\
 & = &  \frac{1}{\eta}\int_\omega d\lambda^1\,d\lambda^2\left(\vec S^1.\delta \vec h_1 + \vec
S^2.\delta \vec h_2 \right)\,,\mylab{fvar}
\end{eqnarray}
where the stress vectors $\vec S^1$ and $\vec S^2$ are given by
\begin{eqnarray}
\vec{S}^{1} & = &
\frac{N^{1}}{l^1}\vec{B}\,^{1}-\frac{N^{3}}{l^3}\vec{B}\,^{3}+%\nonumber\\
% & & 
\vec M^{13}\wedge\left[\frac{\vec B^1}{(l^1)^2}+\frac{\vec B^3}{(l^3)^2}\right]\mylab{s1}\\
\vec{S}^{2} & =& \frac{N^{1}}{2l^1}\vec{B}\,^{1}+\frac{N^{2}}{l^2}\vec{B}\,^{2}+\frac{N^{3}}{2l^3}\vec{B}\,^{3}+%\nonumber\\
%& & 
\frac{1}{2}\vec M ^{13}\wedge\left[\frac{\vec B^1}{(l^1)^2}-\frac{\vec B^3}{(l^3)^2}\right]\mylab{s2}
\end{eqnarray}
Integrating Eq.\,(\ref{fvar}) by parts and taking into account the
periodicity in $\lambda^2$ (i.e.the integrals along the boundaries of
the network with $\lambda^2=0$ and $\lambda^2=2\pi$ cancel each other out)
one finds
\begin{eqnarray}
\delta F & =& -\frac{1}{\eta}\int_\omega \left(\partial_1 \vec S^1 + \partial_2\vec
S^2\right).\delta\vec\varphi\,d\lambda^1\,d\lambda^2 \nonumber\\
& & +\frac{1}{\eta}\int_0^{2\pi} \vec S^1(\alpha,\lambda^2).\delta\vec\varphi\,d\lambda^2\mylab{ibp}%\\
%& & 
- \frac{1}{\eta}\int_0^{2\pi} \vec S^1(0,\lambda^2).\delta\vec\varphi\,d\lambda^2%\,.\nonumber
\end{eqnarray}
The last two integrals over the boundary at $\lambda^1=0$ and
  $\lambda^1=\alpha$ in Eq.\,(\ref{ibp}) evaluate to zero since
$\vec\varphi$ is fixed at the internal boundary,
i.e. $\delta\vec\varphi=0$ for $\lambda^1=0$, and there are no normal
forces acting on the external surface, i.e. $\boldsymbol{\sigma}
n=\vec S^1/\|\vec h^2\|=0$ for $\lambda^1=\alpha$. Here
$\boldsymbol{\sigma}$ denotes the Cauchy stress tensor (see {\bf 
\ref{appb}}) and $\vec n$ is the unit outward surface normal vector in
the deformed configuration.

The mechanical equilibrium in the absence of volumic forces is
therefore given by
\begin{equation}
0=\partial_1\vec S^1 + \partial_2 \vec S^2\mylab{equ}
\end{equation}
with the boundary conditions as mentioned before
\begin{eqnarray}
\forall \lambda^1 & \in & ]0,\alpha[\,,\,
\vec\varphi(\lambda^1,0)  = \vec\varphi(\lambda^1,2\pi)\nonumber\\
\forall \lambda^2 & \in &  ]0,2\pi[\,,\, \vec\varphi(0,\lambda^2) =  R_0\vec e_r(\lambda^2)\mylab{bc}\\
\forall \lambda^2 & \in & ]0,2\pi[\,,\, \vec S^1(\alpha,\lambda^2) = 0\nonumber
\end{eqnarray}
where $\vec e_r=\cos{\lambda^2}\vec e_1+\sin{\lambda^2}\vec e_2$
denotes the unit vector in the radial direction in the orthonormal
coordinate basis $\{\vec e_1,\vec e_2\}$.
%`
For a later use we introduce here also the tangential unit vector as $\vec
e_\theta=-\sin\lambda^2\vec e_1 + \cos\lambda^2\vec e_2$.

\section{Results}
\mylab{results}
\subsection{Axisymmetric state}
\mylab{symres} In the following we will use the simplified notation
$\varphi_r'=\partial_1\varphi_r$.
Eq.\,(\ref{equ}) with the boundary conditions (\ref{bc}) was analyzed
for axisymmetric solutions of the type $\vec
\varphi=\varphi_r(\lambda^1)\vec e_r$.
The stress vectors (\ref{s1}) and (\ref{s2}) can then be simplified to
\begin{eqnarray}
\vec S^1 & = & \eta\left\{2\frac{N^1}{l^1}\varphi_r'-M^{13}\varphi_r^2\varphi_r'\right\}\vec e_r\\
\vec S^2 & = & \eta\left\{\frac{N^1}{2l^1}\varphi_r+\frac{N^2}{l^2}\varphi_r +M^{13}\varphi_r\varphi_r'^2\right\}\vec e_\theta
\end{eqnarray}
The equilibrium equation (\ref{equ}) becomes
\begin{equation}
0  = \partial_1\left(2\frac{N^1}{l^1}\varphi'_r-M^{13}\varphi_r^2\varphi_r'\right)-
\left(\frac{1}{2}\frac{N^1}{l^1}\varphi_r+\frac{N^2}{l^2}\varphi_r+M^{13}\varphi_r\varphi_r'^2\right)\mylab{equ0}
\end{equation}
with the boundary conditions
\begin{eqnarray}
\varphi_r(0) & =& R_0 \nonumber\\
0 & = & \left[2\frac{N^1}{l^1}\varphi_r'-M ^{13}\varphi_r^2\varphi_r'\right]_{\lambda^1=\alpha}\,.\mylab{bc0}
\end{eqnarray}
Eq.\,(\ref{equ0}) was solved using continuation techniques
\cite{AUTO97} and the solutions were characterized using the
equilibrium network thickness $H=\varphi_r(\alpha)-R_0$ and the
associated free energy (\ref{elfe}).

At this stage we point out that the unknown function
$\varphi_r(\lambda^1)$ needs only to be piecewise continuously
differentiable. It can have ``corner points'', i.e.~discontinuities in
the first order derivative.
The presence of such discontinuous solutions
can be inferred from the fact that $\partial^2 f/\partial
\varphi_r'^2$ vanishes for a finite value $\varphi_r'$ at the ``corner
point'' $(\lambda^1_d,\varphi_{rd})$.
For every continuously differentiable segment of an extremal solution
Eq.~(\ref{equ0}) holds supplemented by the following boundary
conditions at the ``free corner points'' \cite{MathHBook}
\begin{eqnarray}
[\partial f/\partial \varphi'_r]_{\lambda^1 = \lambda^1_d-0}& = &
[\partial f/\partial \varphi'_r]_{\lambda^1=\lambda^1_d+0}\mylab{ew1}\\
\left[f-\varphi'_r \partial f/\partial \varphi'_r\right]_{\lambda^1 =
\lambda^1_d-0}& = & \left[f-\varphi'_r \partial f/\partial
\varphi'_r\right]_{\lambda^1 = \lambda^1_d+0}\mylab{ew2}
\end{eqnarray}
The first condition (\ref{ew1}) is equivalent to the physical boundary
condition, that the normal stresses match at the ``cornerpoint''.
The second condition (\ref{ew2}) minimizes the energy with respect to
the position of the ``corner point''.

%
%
% special case $k^2=0$
%
First one might consider the special case $k^2=\kappa=0$. As it is
easy to see, $N^1=0$ solves Eqs.\,(\ref{equ0}) and (\ref{bc0}). Under
the condition that $l_m^1\geq p/2$ (recall that $p$ denotes the
distance at which the nodes are attached at the cylinder surface) one
therefore finds the equilibrium solution
\begin{equation}
\frac{\varphi_r}{R_0}=\frac{2 l_m^1}{p}
\sin{\left[\frac{\lambda^1}{2}+\arcsin{\left(\frac{p}{2
l_m^1}\right)}\right]}\,.\mylab{equsol}
\end{equation}
At the radial layer number
$\lambda^1=\lambda^1_c=\pi-2\arcsin{\left(\frac{p}{2l_m^1}\right)}$
one finds $\varphi_r'=0$ and $\varphi_r/R_0=2l_m^1/p$. At this point
the network collapses to zero volume and the filaments $\vec B^1$,
$\vec B^2$ and $\vec B^3$ are oriented in parallel.
The addition of further network layers will either lead to a
back-folding of the gel into itself (in this case solution
(\ref{equsol}) holds), or the network layer will rest in this
collapsed state, i.e.~$\varphi_r/R_0=2l^1_m/p$ for
$\lambda^1>\lambda^1_c$. In this case $\lambda^1_c$ constitutes
a ``corner point'', where Eqs.~(\ref{ew1}) and (\ref{ew2}) hold.

The back-folding behavior is excluded in a physical situation since it
requires the remodeling of the network structure and in the following
we will not consider these special cases. The network collapse is
unphysical as well since it implies a zero volume of the network
layers.

Now we will consider the more physical case $k^2>0$.  Fig.\,\ref{fk2}
shows the dependence of the equilibrium network thickness $H$ and the
associated energy depending on the number of radial network layers
$\alpha$ for the case $k=k^2/k^1=1$ and $l_m^1=l_m^2=p$, i.e.\, all
filaments are identical and type~2 filaments are inserted at the
internal interface at their equilibrium length.
\begin{figure}[hbt]
\begin{center}
\includegraphics[width=0.7\hsize]{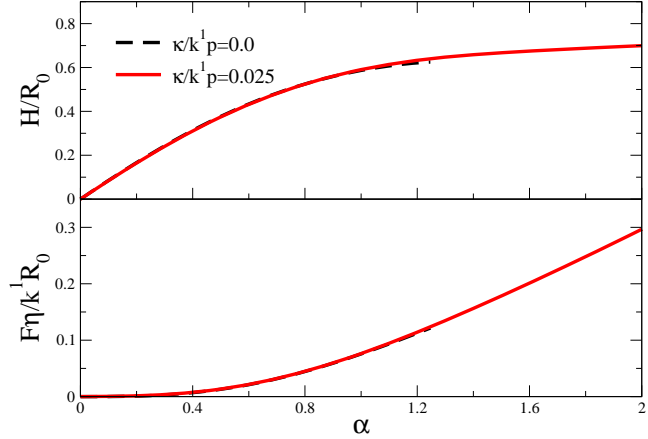}
\end{center}
\caption{[Color online] (a) Equilibrium thickness of the network $H$
  and (b) the associated energy depending on the number of radial
  network layers $\alpha$ for $\kappa=0$ and $\kappa/k^1p=0.025$ as
  indicated in the legend. Remaining parameters are $k_2/k_1=1$ and
  $l^1_m=l^2_m=p$.\mylab{fk2}}
\end{figure}
First, increasing the ratio of the elastic constant $k^2/k^1$
decreases the network thickness $H$ for a constant number of radial
network layers $\alpha$.  Second, the continuously differentiable
solution branches with $\varphi_r'>0$ cease to exist beyond a critical
value $\alpha_c$.  Instead solutions appear which have one or more
discontinuities in $\varphi_r'$ and which contain solution segments
with $\varphi_r'<0$ (not shown).

To exclude the back-folding behavior we have added an additional term
in the free energy related to the interaction between $\vec B^1$ and
$\vec B^3$ and which prevents the alignment of $\vec B^1$ and $\vec
B^3$. Fig.\,\ref{fk2} shows exemplary the network thickness and free
energy for a small value of $\kappa/k^1$.  Solutions exist even for
very large networks and the solution behavior for small networks is
similar to the behavior with $\kappa=0$.

Fig.\,\ref{sol1} shows exemplary solutions for
Eq.\,(\ref{equ0}) for small, intermediate and large networks.
For small and intermediate networks, the filament lengths and stresses
increase almost linearly with $\lambda^1$, whereas for large networks
the behavior is strongly nonlinear. The network is always under radial
compression, whereas it is under tangential compression close to the
surface of the obstacle and under tangential extension at the outer
surface of the network. 
\begin{figure}[hbt]
\begin{center}
\includegraphics[width=0.7\hsize]{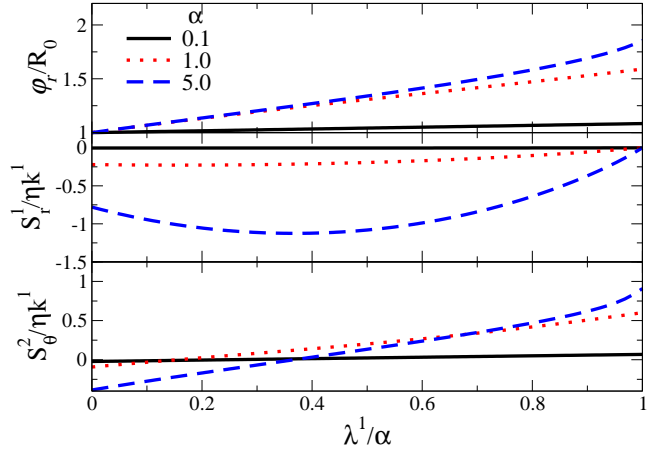}
\end{center}
\caption{[Color online] Solutions of Eq.\,\ref{equ0} for different
  network thicknesses $\alpha$ as indicated in the legend. Shown is
  (a) the radial node position, (b) the radial stress, and (c) the
  tangential stress distribution. Remaining parameters are
  $k^2/k^1=1$, $\kappa/k^1p=0.025$, $l_m^1=l_m^2=p$.\mylab{sol1}}
\end{figure}
%

%
% linear stability
%
\subsection{Linear stability}
\mylab{linstab}
To study the (mechanical) linear stability of the axisymmetric
state we have evaluated the sign of the second variation of $F$ with
respect to $\vec \varphi$ at the prestress axisymmetric state
$\varphi_r \vec e_r$
\begin{equation}
\delta^2 F=\int_\omega \delta\vec h_i\frac{\partial^2f}{\partial\vec h_i \partial\vec h_j} \delta\vec h_j d\lambda^1\,d\lambda^2
\mylab{2nd}
\end{equation}
with the boundary conditions 
\begin{eqnarray}
\forall \lambda^1 & \in & ]0,\alpha[\,,\,\delta\vec \varphi(\lambda^1,0)= \delta\vec \varphi(\lambda^1,2\pi)\nonumber\\
\forall \lambda^2 & \in &  ]0,2\pi[\,,\, \delta\vec\varphi (0,\lambda^2) =  0 \mylab{bc1}
\end{eqnarray}
Details of the stability analysis can be found in {\bf
  \ref{applinstab}}. Briefly, we have used a Fourier mode
decomposition of $\delta \vec \varphi$ and a direct integration in
$\lambda^2$ and finite elements for a numerical integration in the
$\lambda^1$.

Fig.\,\ref{phase} shows the linear stability of $\varphi_r(\lambda^1)$
depending on the elastic ratio $k^2/k^1$ and $\alpha$.
\begin{figure}
\begin{center}
\includegraphics[width=0.7\hsize]{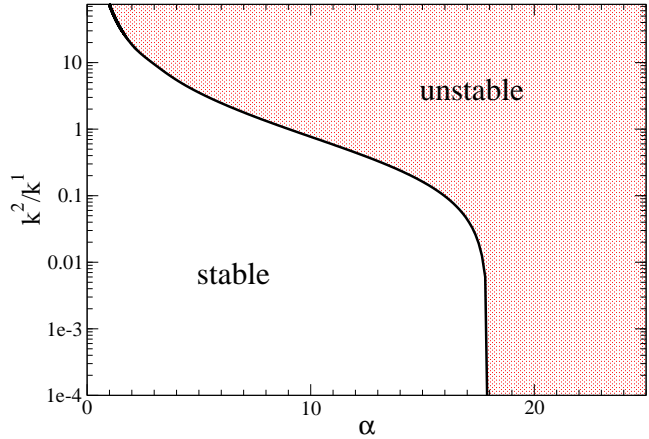}
\end{center}
\caption{Linear stability of continuously differentiable axisymmetric
solutions depending on the parameters $k^2/k^1$ and $\alpha$. Above
(below) the solid line axisymmetric solutions are linearly unstable
(stable). Remaining parameters are $l_m^1=l_m^2=p$ and
$\kappa/k^1p=0.025$. \mylab{phase}}
\end{figure}
The linear stability of the axisymmetric state depends on
the number of radial network layers superimposed over each other,
whereby thin networks are stable and thick networks are
unstable. Furthermore, networks with higher tangential stresses (large
$k^2/k^1$) destabilize at smaller $\alpha$.
The instability occurs independently of the transversal wavenumber $m$
and is accompanied by the appearance of a region adjacent to the inner
interface where $\partial^2 f/\partial h_{1\theta}^2<0$. 
The appearance of the instability is probably due to the fact, that as
the network thickness increases the compressive load acting on type 1
and type 3 filaments in the innermost network layers increases. The
elastic stress is released by rotating consecutive network layers
against each other. 
Since the model does not contain any force opposed to a change in
orientation of consecutive filaments of type 1 and 3, which
introduce an additional length scale related to a ``persistence
length'' of consecutive filaments of the same type, a catastrophic
growth of the instability at infinitely small scales occurs (data not
shown).

\section{Conclusions}
We have introduced a two-dimensional homogenization model to describe
the mechanics of stiff elastic filament networks assembled on a
cylindrical surface resembling Arp2/3 cross-linked actin networks as
they occur in living cells.
Our approach allows us to realistically model macroscopic deformations and
stresses due to network growth by using only assumptions about the
microscopic network properties.
The main purpose of the model is to introduce a minimal set of
mechanical ingredients as a basis for the more advanced problem of the
time-dependent evolution of the network interfaces due to
polymerization/depolymerization at the interface with a possible
symmetry-breaking and subsequent motion.

The approach is novel in the field of actin driven motion, inasmuch as
it allows to calculate equilibrium stresses and the position of the
interface between the network and the surrounding solution without any
ad hoc assumptions \cite{SPJ04} or restrictions to infinitesimal
deformations \cite{JPK08}. Furthermore it will allow a rigorous
calculation of the equilibrium state for networks with perturbed
interfaces, resulting from spontaneous variations in the
polymerization or depolymerization speed at the network interfaces as
outlined in \cite{JCP09}.
However, in this study we have limited ourselves to fixed interfaces
(in the Lagrangian frame) and studied the existence and linear
stability of physical and biological relevant axisymmetric equilibrium
states.

We find that in a triangular network where only changes in the length
of the edges contribute to the free energy the existence of
sensible solutions is limited by the number of radial network layers
$\alpha$ superimposed over each other. As one increases $\alpha$
beyond a critical number the network starts to fold back into the
negative radial direction, since the extension of the microscopic
filaments in the tangential direction is unfavorable as one moves
away from the cylinder surface. The inclusion of an additional term in the
free energy related to the interaction between two selected filaments
prevented this back-folding.

Furthermore, the linear stability of the axisymmetric solution depends
on the microscopic elastic properties. In general, thin networks are
stable and thick networks are unstable, whereby a higher elastic
constant for filaments oriented in the tangential direction favors the
occurrence of the instability for thinner networks.
The instability is independent of the transversal wavenumber of the
harmonic perturbations. A closer inspection of the instability reveals
that it occurs first at the cylinder surface and is reminiscent of a
buckling instability of type 1 and 3 filament, i.e. a counterrotation
of consecutive radial network layers.
However, there is no mechanism in this simple model to stabilize this
instability. A more advanced approach would have to include a more
detailed microscopic mechanical model with nonlocal contributions
(i.e. in the spirit of second gradient model), which is beyond the scope of this paper and
will be the subject of further investigation.

%% The Appendices part is started with the command \appendix;
%% appendix sections are then done as normal sections
\appendix

\section{Derivation of the Cauchy stress tensor in the network}
\mylab{appa}
The first to define a stress tensor in a discrete medium in the
Lagrangian frame were Cauchy and Poisson whose derivation can be found
in \cite{Lov44,CaC01}. We present here the basic ideas in
two dimensions to facilitate the comprehension of the {\bf Theory}
section.

For the derivation of the stress one considers a periodic structure
of nodes and their connections. Forces act only along the directions
of the node connections. The periodic structure is divided into
elementary cells which are numbered by $(\nu^1,\nu^2)$. The
connections are identified by the triplet $(b,\nu^1,\nu^2)$, where $b$
identifies the connection within the elementary cell.
%
%The unit vectors along the node connections are denoted by $\vec e^b$
%and their length is $l^b$.
%

\begin{figure}
\begin{center}
\includegraphics[width=0.7\hsize]{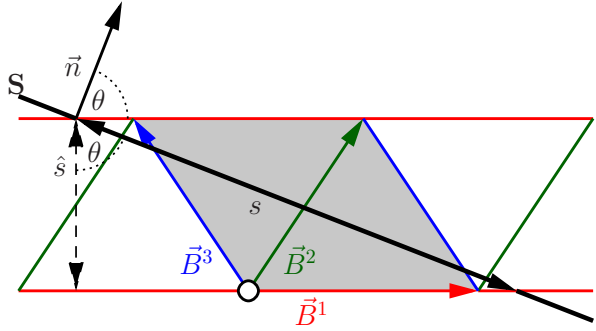}
\end{center}
\caption{A facet $S$ with the unit normal $\vec n$ and the segment
length $s$ is placed in the network. The elementary cell with the
basis vectors $\vec B^b$ and the height $\hat s$ is shown in
gray.\mylab{facet}}
\end{figure}
The stress tensor is defined over a force $\vec R$ acting on a
arbitrary facet S with unit normal $\vec n$ placed in the periodic
medium as shown in Fig.\,\ref{facet}.
The force $\vec R$ is obtained by summing up the forces acting along
the node connections which are cut by the facet. Thereby one assumes
that the length $S$ of the facet is sufficiently small that the forces
exerted by identical connections are constant over the facet.
Therefore the force $\vec R$ is given by
\begin{equation}
\vec R=\sum_b m^b \mathrm{sgn}(\vec B^b.\vec n) N^b \frac{\vec B\,^b}{\|\vec B^b\|}\,.\mylab{vecR}
\end{equation}
Here $N^b$ and $m^b$ denote the tension and number of node connections
of type $b$. The term $\mathrm{sgn}(\vec B^b.\vec n)$ accounts for the
orientation of the connections.
There are different ways to determine $m^b$, we give here only the
result derived in detail in \cite{CaC01}.
The basic idea is, that the facet $S$ is cut by the vectors of one
type $b$ into segments $s$ and therefore $m^b=S/s$. As can be
seen in Fig.\,\ref{facet} for the example of the type~1 filament, the
height of the elementary cell is $\hat s=s|\vec B^1.\vec n|/\|\vec
B^1\|$ and consequently its area is given by $A_c=\hat s \|\vec
B^1\|$.  Therefore the number of links of type $b$ cut by the facet S
can be expressed as
\begin{equation}
m^b = \frac{S\,| \vec B\,^b.\vec n|}{A_c} \,\mylab{number}
\end{equation}
Using expression
(\ref{number}) with Eq.\,(\ref{vecR}) one obtains
\begin{equation}
\vec R =\frac{1}{A_c}\sum_b N^b \frac{\vec B\,^b}{\|\vec B^b\|}(\vec B\,^b.\vec n) S\,.
\end{equation}
After some rearrangements and using the identity
\begin{equation}
\vec R=\boldsymbol{\sigma} \vec n S
\end{equation}
one finds
\begin{equation}
\boldsymbol{\sigma}=\frac{1}{A_c}\sum_b N^b \frac{\vec B\,^b \otimes \vec B\,^b}{\|\vec B^b\|}\mylab{sigcau}
\end{equation}
%
%
%----------------------------------------------------------------------%
%
\section{Relation between the stress vectors $\vec S^i$ and the Cauchy
  stress tensor $\boldsymbol{\sigma}$}
\mylab{appb}
In the present section we demonstrate the relation between the
classical equilibrium equation known from continuum mechanics using
the Cauchy stress tensor $\boldsymbol{\sigma}$ in the deformed
configuration and the equilibrium formulation we have used throughout
this paper defined on the Lagrangian reference configuration.
Note, that we will use the Einstein notation, i.e.\,summation over
repeated indices, and that we will use the abbreviation
$\partial_i=\partial_{\lambda^i}$ which denotes a partial derivation
with respect to the coordinate $\lambda^i$ in the Lagrangian reference
configuration.

We start out by introducing some equalities.
Let $\vec \varphi$ be the deformation mapping of the network with
respect to its Lagrangian configuration $\omega$ and let $\vec x=\vec
\varphi \left( \lambda^1,\lambda^2 \right)$ be the position in the
deformed configuration $\omega_\varphi =\vec \varphi
\left(\omega\right)$ of the network point $(\lambda^1,\lambda^2)$ of
the Lagrangian reference configuration. Let $d\vec x$ be the
difference of positions of two neighbouring points $(
\lambda^1,\lambda^2)$ and $( \lambda^1+d \lambda^1,\lambda^2+d
\lambda^2)$, then
\begin{equation} \label{eq:dx}
d\vec x = \vec h_i d\lambda^i
\end{equation}
where $\vec h_i=\partial_i\vec \varphi$. Consequently one finds
\begin{equation} \label{eq:dlambdai}
d\lambda^i=\vec h^i\ldot d\vec x
\end{equation}
with the property
\begin{equation}
\vec h_i \ldot\vec h^j= \delta^j_i \mylab{contrabase}
\end{equation}

Let now $\vec v$ be a virtual velocity field  which, due to the one to one mapping $\vec \varphi$ from $\omega$ to $\omega_\varphi$, can be considered either as a function of $\lambda^1,\lambda^2$ or as a function of $\vec x$. $\vec v$ is assumed to be differentable and let $\nabla \vec v$ be its differential with respect to $\vec x$ and $\partial_i \vec v$ be its partial derivative with respect to $\lambda^i$, then
\begin{align}
d\vec v&=\nabla \vec v d\vec x \\
d\vec v&=\partial_i \vec v d\lambda^i
\end{align}

Now, the use of equations (\ref{eq:dx}) and  (\ref{eq:dlambdai}) yields
\begin{align}
\nabla \vec v =\partial_i \vec v\otimes \vec h^i \\
\partial_i \vec v =\nabla \vec v\,\vec h_i 
\end{align}

In the absence of body forces the virtual power formulation of the equilibrium equation (\ref{equ}) reads
\begin{equation}
\int_\omega\vec S^i. \partial_i \vec v \, d\lambda^1d\lambda^2=0
\end{equation}
where the virtual velocity field $\vec v$ vanishes on the boundaries $\lambda^1=0$ and $\lambda^1=\alpha$ of $\omega$ and is $2\pi\,\mathrm{periodic}$ with respect to $\lambda^2$.

The change of variables $(\lambda^1,\lambda^2)\leftrightarrow \vec x=\varphi(\lambda^1,\lambda^2)$ in the intergral of the previous equation yields
\begin{equation}
\int_{\omega_\varphi}\vec S^i.( \nabla \vec v\,\vec h_i)\,  \frac{1}{h}ds=0
\end{equation}
where $ds=h\, d\lambda^1 d\lambda^2$ denotes the area element of the body in the deformed configuration with $h=\|\vec h_1 \wedge \vec h_2\|$.
The previous equation also reads
\begin{equation}
\int_{\omega_\varphi}  \frac{1}{h}\left( \vec S^i\otimes \vec h^i \right): \nabla \vec v\, ds=0
\end{equation}
where $\mathbf{A}:\mathbf{B}$ denotes the usual twice contracted product between the second order tensors $\mathbf{A}$ and $\mathbf{B}$.

Comparing with the classical virtual power formulation of the equilibrium in the Eulerian variable $\vec x$ which reads
\begin{equation}
\int_{\omega_\varphi} \boldsymbol{\sigma}: \nabla \vec v\, ds=0
\end{equation}
we can see that the expression of the Cauchy stres tensor $\boldsymbol{\sigma}$ in terms of the $\vec S^i$'s reads
\begin{equation}
\boldsymbol{\sigma}=\frac{1}{h}\vec S^i\otimes\vec h_i \mylab{sigs}
\end{equation}

Conversely, the use of definition (\ref{contrabase}) of the contravariant base $\left( \vec h^1,\vec h^2 \right)$ in the equation (\ref{sigs}) yields the expressions of the vectors $\vec S^1$ and $\vec S^2$ in terms of $\boldsymbol{\sigma}$ which reads
\begin{equation}
\vec S^i=h\boldsymbol{\sigma}\vec h^i
\end{equation}

\section{Linear stability analysis}
\mylab{applinstab}
Since the boundary conditions are periodic in $\lambda^2$ one can
decompose $\delta\vec
\varphi= \delta\varphi_r\vec e_r+\delta\varphi_\theta\vec e_\theta$ into a Fourier
series
\begin{eqnarray}
\delta \varphi_r  & = &  \sum_{m=0}^\infty \left[\delta\varphi^m_{rc}(\lambda^1)\cos{(m\lambda^2)}+ \delta\varphi^m_{rs}(\lambda^1)\sin{(m\lambda^2)}\right] \nonumber\\
\delta \varphi_\theta  & = &  \sum_{m=0}^\infty \left[\delta\varphi^m_{\theta c}(\lambda^1)\cos{(m\lambda^2)}+ \delta\varphi^m_{\theta s}(\lambda^1)\sin{(m\lambda^2)}\right] \nonumber\\
\mylab{fsd}
\end{eqnarray}
where $m$ denotes the wavenumber and $\delta\varphi_{rc}^m$,
$\delta\varphi_{rs}^m$, $\delta\varphi_{\theta c}^m$,
$\delta\varphi_{\theta s}^m$ denote perturbation amplitudes which
depend only on $\lambda^1$. Using ansatz (\ref{fsd}) one
can perform the integration of (\ref{2nd}) in $\lambda^2$ directly
whereby different wavenumbers $m$ decouple. One finds then
\begin{equation}
\delta^2 F = \pi\sum_{m=0}^\infty\int_0^\alpha \left(\delta^2f^m_1+\delta^2f^m_2 \right)d\lambda^1\mylab{2nd1}
\end{equation}
where $\delta^2f^m_{1,2}$ takes the form
\begin{eqnarray}
\delta^2 f^m_{1,2}& =&A(a')^2+B(b')^2+2C(a'a+ma'b)-\\
&  &2D(b'b+mab')+E(ma+b)^2+F(a+mb)^2\nonumber
\end{eqnarray}
with $(a,b)=(\delta\varphi_{rc}^m,\delta\varphi_{\theta s}^m)$ and
$(a,b)=(-\delta\varphi_{rs}^m,\delta\varphi_{\theta c}^m)$ for $\delta^2 f^m_1$ and
$\delta^2 f^m_2$, respectively.
The factors $A,\ldots, F$ are evaluated at the axisymmetric solution
$\vec\varphi=\varphi_r\vec e_r$ and are given by 
\begin{eqnarray}
A & = & \frac{\partial^2 f}{\partial h_{1r}^2}=2\frac{N^1}{l^1}+2k^1\frac{\varphi_r'^2}{(l^1)^3}+\kappa\frac{3\varphi_r^4/16-\varphi_r'^4}{\varphi_r^2\varphi_r'^4}\nonumber\\
B & = & \frac{\partial^2 f}{\partial h_{1\theta}^2}=2\frac{N^1}{l^1}+k^1\frac{\varphi_r^2}{2(l^1)^3}-2\kappa\frac{\varphi_r^2/4-\varphi_r'^2}{\varphi_r^2\varphi_r'^2}\nonumber\\
C & = & \frac{\partial^2 f}{\partial h_{1r}\partial h_{2\theta}} = k^1\frac{\varphi_r\varphi_r'}{2 (l^1)^3} -2\kappa\frac{\varphi_r^4/16+\varphi_r'^4}{\varphi_r^3\varphi_r'^3}\nonumber\\
D& = & \frac{\partial^2 f}{\partial h_{1\theta}\partial h_{2r}} = k^1\frac{\varphi_r\varphi_r'}{2 (l^1)^3}+\kappa\frac{(\varphi_r^2/4-\varphi_r'^2)^2}{\varphi_r^3\varphi_r'^3}\nonumber\\
E & = & \frac{\partial^2 f}{\partial h_{2r}^2}=\frac{N^1}{2l^1}+\frac{N^2}{l^2}+k^1\frac{\varphi_r'^2}{2(l^1)^3}+\kappa\frac{\varphi_r^2/4-\varphi_r'^2}{2\varphi_r^2\varphi_r'^2}\nonumber\\
F& = & \frac{\partial^2 f}{\partial h_{2\theta}^2}=\frac{N^1}{2l^1}+\frac{N^2}{l^2}+k^1\frac{\varphi_r^2}{8(l^1)^3}+k^2\frac{\varphi_r^2}{(l^2)^3}+\nonumber\\
& & \kappa\frac{\varphi_r^4/16+3\varphi_r'^4}{\varphi_r^4\varphi_r'^2}\nonumber
\end{eqnarray}
The integral (\ref{2nd1}) was calculated numerically by using finite
elements, whereby the functions $\delta\varphi_{rs}^m$,
$\delta\varphi_{rc}^m$, $\delta\varphi_{\theta s}^m$, and
$\delta\varphi_{\theta c}^m$ were approximated by P$_1$-elements and
the functions $A,\ldots, F$ were approximated by P$_0$ elements. This
results in the quadratic form
\begin{equation}
\delta^2 F =  \sum_{m=0}^\infty\left\{(\hat\varphi_{rc}^m,\hat\varphi_{\theta s}^m) \mathbf{Q}^m (\hat\varphi_{rc}^m,\hat\varphi_{\theta s}^m)^T+%\right.\nonumber\\
% &  &\left.
(-\hat\varphi_{rs}^m,\hat\varphi_{\theta c}^m) \mathbf{Q}^m (-\hat\varphi_{rs}^m,\hat\varphi_{\theta c}^m)^T\right\}
\end{equation}
where $\hat{}$ indicates a vector containing function values evaluated
at the discretization points $\lambda^1_i, (i=1,2,\ldots,n)$ and
$\mathbf{Q}^m$ is a symmetric $2n\times 2n$ matrix.

For the axisymmetric state to be linearly stable $\delta^2F>0$,
i.e.~$\mathbf{Q}^m$ needs to be a positive definite matrix. We have
tested this condition for wavenumbers $m\le 4$. But it turned out that
an instability appears independently of $m$ when the factor
$B=\partial^2f/\partial h^2_{1\theta}$ changes sign at
$\lambda^1=0$. The corresponding phase diagram is shown in
Fig.\,\ref{phase}.

%% \section{}
%% \label{}

%% References
%%
%% Following citation commands can be used in the body text:
%% Usage of \cite is as follows:
%%   \cite{key}         ==>>  [#]
%%   \cite[chap. 2]{key} ==>> [#, chap. 2]
%%

%% References with BibTeX database:

\bibliographystyle{elsarticle-num}
%\bibliography{../Bibs/elasticity%
%,../Bibs/eigene%
%,../Bibs/motility%
%}

%% Authors are advised to use a BibTeX database file for their reference list.
%% The provided style file elsarticle-num.bst formats references in the required Procedia style

%% For references without a BibTeX database:

% \begin{thebibliography}{00}

%% \bibitem must have the following form:
%%   \bibitem{key}...
%%

% \bibitem{}

% \end{thebibliography}

\end{document}